\documentclass[preprint,aps,superscriptaddress,showpacs,floatfix]{revtex4}\usepackage{amssymb}
\usepackage{amsmath}
\usepackage{graphicx}
\usepackage[normalem]{ulem}
\usepackage[dvips]{color}
\usepackage{bm}
\usepackage{longtable}

\renewcommand\sout{\bgroup \color{red} \ULdepth=-.5ex \ULset}

\begin{document}

\title{Effects of short-range correlation reduced kinetic symmetry energy in heavy-ion collisions at intermediate energies}

\author{Bao-An Li}
\affiliation{Department of Physics and Astronomy, Texas A$\&$M University-Commerce, Commerce, TX 75429-3011, USA}
\author{Wen-Jun Guo}
\affiliation{Department of Physics and Astronomy, Texas A$\&$M University-Commerce, Commerce, TX 75429-3011, USA}
\affiliation{College of Science, University of Shanghai for Science and Technology, Shanghai, 200093, China}
\author{Zhaozhong Shi}
\affiliation{Department of Physics and Astronomy, Texas A$\&$M University-Commerce, Commerce, TX 75429-3011, USA}
\affiliation{University of California at Berkeley, Berkeley, CA 94720, USA}


\begin{abstract}
Besides earlier predictions based on both phenomenological models and modern microscopic many-body theories, circumstantial evidence was recently found for a reduced kinetic symmetry energy of isospin-asymmetric nucleonic matter compared to the free Fermi gas model prediction due to the short-range correlation of high-momentum neutron-proton pairs. While keeping the total symmetry energy near the saturation density of nuclear matter consistent with existing experimental constraints, we examine the correspondingly enhanced role of the isospin degree of freedom in heavy-ion collisions at intermediate energies due to the reduced (enhanced) kinetic (potential) symmetry energy. Important observable consequences are investigated.
\end{abstract}

\maketitle

\section{Introduction}
To pin down the isospin-dependent term of the Equation of State (EOS) of neutron-rich nucleonic matter, i.e., the density $\rho$ dependence of nuclear symmetry energy $E_{sym}(\rho)$,  is a common goal of many studies in both nuclear physics and astrophysics, see, e.g., ref.\ \cite{EPJA} for a recent and comprehensive  review. To  achieve this goal, it is important to know more about the origin of the symmetry energy. The symmetry energy has kinetic and potential parts.  In many studies, the kinetic symmetry energy is normally approximated by a free Fermi gas model prediction
\begin{equation}\label{FG}
E_{sym}^{kin}(\textrm{FG})(\rho)\equiv(2^{\frac{2}{3}}-1)\frac{3}{5}E_F(\rho)\approx 12.5(\rho/\rho_0)^{2/3}
\end{equation}
where $E_F(\rho)$ is the Fermi energy at density $\rho$. However, this approximation was recently found to be invalid when effects of the isospin-dependent short-range nucleon-nucleon correlations are considered. In particular, it was shown in both phenomenological models \cite{XuLi} and microscopic many-body theories \cite{Vid11,Lov11,Car12,Rio14} that the short-range correlation (SRC) due to the tensor force acting predominately between a spin-triplet, isospin-singlet neutron-proton pair reduce significantly the kinetic symmetry energy to even negative values at saturation density $\rho_0$. Moreover, circumstantial evidence supporting this prediction was recently found from analyzing both (e,e$^{\prime}$) scattering \cite{Hen14} and heavy-ion collision experiments \cite{OrHen14}.  Since the total symmetry energy at $\rho_0$ is relatively well determined to be around a global average of $S_0\equiv E_{sym}(\rho_0)=31.6\pm 2.66$ MeV \cite{LiBA13}, the magnitude of the potential symmetry energy at $\rho_0$ has to be enhanced proportionally. We notice that in situations where only the total symmetry energy matters, such as, the extraction of symmetry energy and its density slope from analyzing atomic masses, $\alpha$ and $\beta$ decay energies, isobaric analog states and the isoscaling parameters, how the $S_0$ is divided into its kinetic and potential parts has no observable effect. However, it matters in dynamical models where the symmetry potential is a direct input.
For example, in transport model simulations of heavy-ion collisions, the symmetry potential corresponding to a given potential symmetry energy is a direct input. On the other hand, the kinetic symmetry energy does not directly enter transport model simulations but limits the magnitude of the potential symmetry energy through the sum rule  $S_0=E^{kin}_{sym}(\rho_0)+E^{pot}_{kin}(\rho_0)$. The enhanced (reduced) potential (kinetic) symmetry energy is expected to affect the significance of the isospin degree of freedom in heavy-ion collisions.  While extensive studies of the potential symmetry energy over a broad density range using heavy-ion experiments have been carried out, information about the kinetic symmetry energy at $\rho_0$ from $(e,e^{\prime})$ scattering experiments just started appearing \cite{Hen14,OrHen14}. Of course, they are complementary to each other and a complete determination of the density dependence of nuclear symmetry requires better knowledge of both kinetic and potential symmetry energies. In this work, within the IBUU transport model \cite{LCK} we examine quantitatively how the role of isospin degree of freedom might be increased by the enhanced (reduced) potential (kinetic) symmetry energy in heavy-ion collisions at intermediate energies.

The paper is organized as follows. We shall first examine how the existing constraints on the magnitude and slope of the $E_{sym}(\rho)$ at $\rho_0$ may limit its division into kinetic and potential parts. Then, within the IBUU transport model using the option of a momentum-independent potential we examine effects of a reduced (enhanced) kinetic (potential) symmetry energy on (1) the time evolution of the neutron/proton ratio in both the gas ($\rho\leq \rho_0/8$) and liquid ($\rho>\rho_0/8$) regions, (2) the free neutron/proton ratio as a function of nucleon kinetic energy and their beam energy dependence, (3) the mid-rapidity neutron/proton ratio as a function of transverse momentum, and (4) the time evolution of the $\pi^-/\pi^+$ ratio in heavy-ion collisions near the pion production threshold. Finally, we summarize.

 \begin{figure}[htb]
\begin{center}
\includegraphics[width=\linewidth]{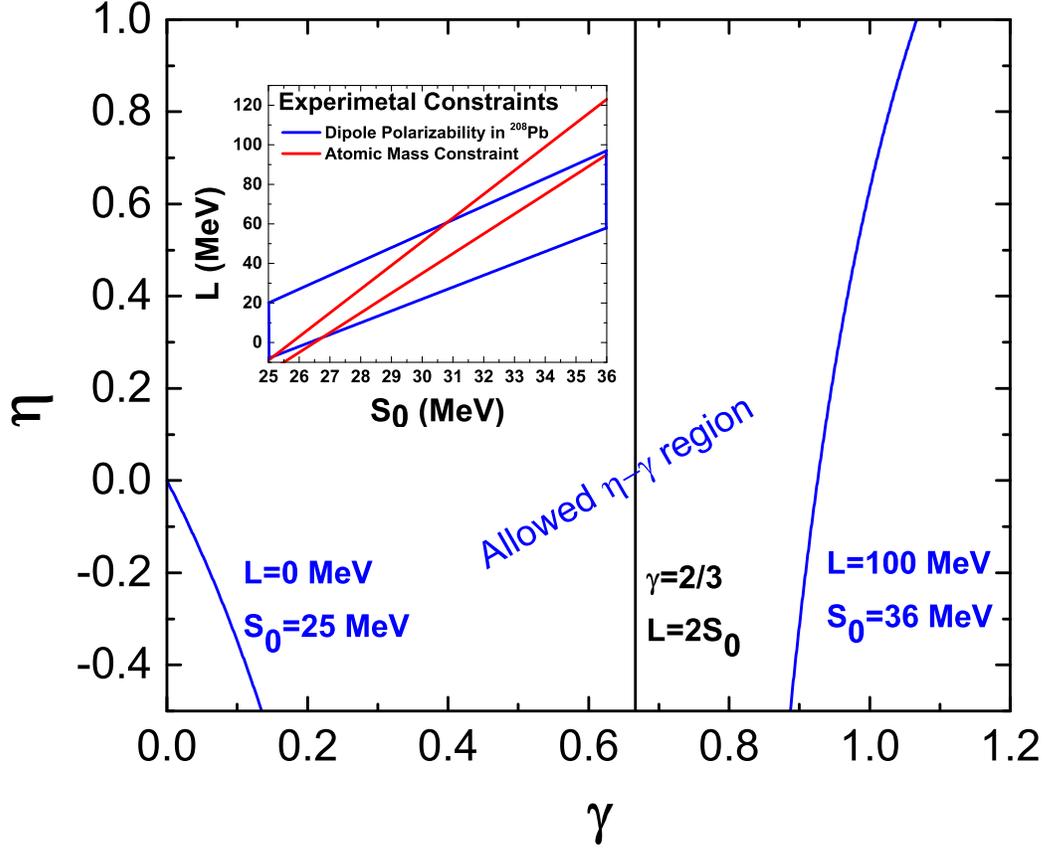}
\caption{(Color online) \label{figure1} The allowed region of the $\eta-\gamma$ plane corresponding approximately to the lower and upper limits of the constraints on the $S_0-L$ correlation shown in the inset.}
\end{center}
\end{figure}

\begin{figure}[htb]
\begin{center}
\includegraphics[width=\linewidth]{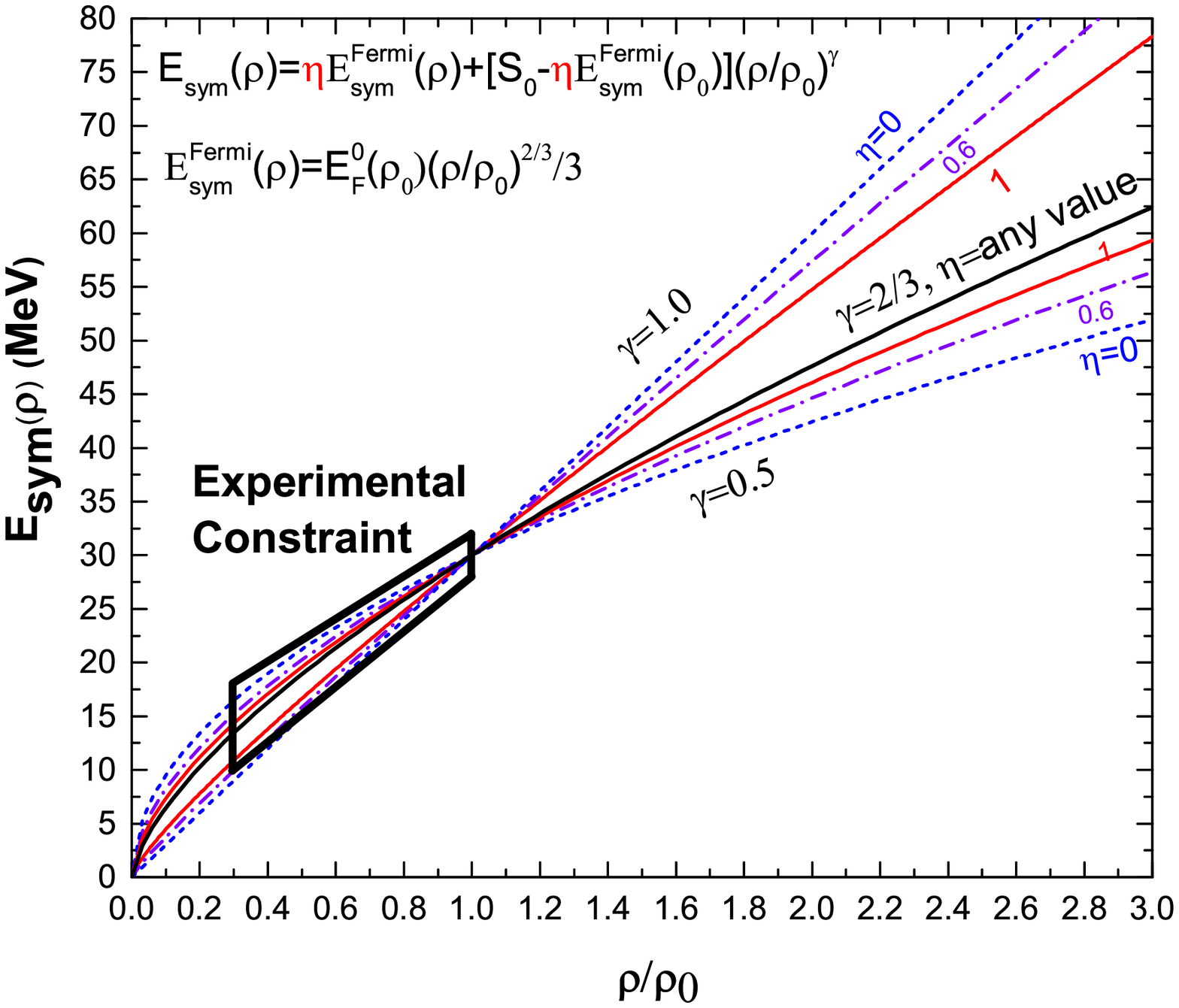}
\caption{(Color online) \label{figure2} The density dependence of nuclear symmetry energy with various combinations of the $\eta-\gamma$ values in comparison with the experimental constraints in the subsaturation density region \cite{Tsang12}.}
\end{center}
\end{figure}

\section{Division of nuclear symmetry energy into its kinetic and potential parts within existing experimental constraints}
Significant progress has been made recently in constraining the density dependence of nuclear symmetry energy around $\rho_0$ \cite{EPJA,Bar05,Lynch09,Trau12,Tsang12,Hor14,Jim13}.
However, the available constraints do not constrain individually the kinetic and potential parts of the symmetry energy. In fact, with the exception of dynamical observables in nuclear reactions, only the total symmetry energy is extracted from model analyses of experimental data.
Assuming the kinetic part is the one given in Eq. \ref{FG},
the potential part $E_{sym}^{pot}(\rho)$ normally contains one or more parameters with its strength limited by the condition $E_{sym}^{pot}(\rho_0)=E_{sym}(\rho_0)-E_{sym}^{kin}(FG)(\rho_0)\approx 19.1$ MeV at $\rho_0$.
Moreover, the correlated Fermi gas model \cite{OrHen14} and the microscopic many-body theories~\cite{Vid11,Lov11,Car12,Rio14} have all indicated consistently that the SRC
reduces the magnitude significantly but affects very little the slope $L\equiv 3\rho(\partial E_{sym}/\partial \rho)_{\rho_0}$ of the kinetic symmetry energy with respect to the free Fermi gas model prediction.  It is thus reasonable to parameterize the symmetry energy as
\begin{equation}\label{esym}
E_{sym}(\rho)=\eta\cdot E_{sym}^{kin}(\textrm{FG})(\rho)+[S_0-\eta\cdot E_{sym}^{kin}(\textrm{FG})(\rho_0)](\frac{\rho}{\rho_0})^{\gamma}
\end{equation}
using two parameters $\eta$ and $\gamma$ to vary its kinetic and potential part, respectively. The corresponding $L$ is
\begin{equation}\label{LL}
L=\frac{9}{5}(2^{2/3}-1)E_F(\rho_0)(2/3-\gamma)\eta+3\gamma S_0.
\end{equation}

At least 30 different analyses so far have attempted to constrain the $S_0-L$ correlation using various data from both terrestrial nuclear laboratory experiments and astrophysical observations.
Shown in the inset of Fig.\ \ref{figure1} are two examples from analyzing atomic masses \cite{lat12} and the dipole polarizability of $^{208}$Pb \cite{Tamii}. Given a set of $S_0$ and $L$, a correlation between $\eta$ and $\gamma$ can be obtained from Eq. \ref{LL}.  Shown in Fig. \ref{figure1} are boundaries in the $\eta-\gamma$ plane between two extremes with $S_0=25$ MeV and $L=0$ on the left and $S_0=36$ MeV and $L=100$ MeV on the right. It is seen that widely diverse combinations of $\eta$ and $\gamma$ are allowed by the existing constraints on the $S_0-L$ correlation. In particular, it is interesting to note from examining Eq. \ref{LL} that if $\gamma=2/3$, then $L=2S_0$ independent of $\eta$, namely any value of $\eta$ is allowed when both the kinetic and potential parts vary with $(\rho/\rho_0)^{2/3}$. Probably incidentally, 28 analyses of various terrestrial and astrophysical data led to the global mean values of $S_0= 31.6\pm 2.6$ MeV and $L=58.9\pm 16.0$ MeV \cite{LiBA13} satisfying approximately the $L=2S_0$ relation. Thus, the currently existing constraints on the $S_0-L$ correlation does not limit even loosely the value of $\eta$, namely the kinetic symmetry energy essentially can be anything.

The range of the allowed $\eta-\gamma$ combinations can also be examined by comparing the symmetry energy parameterized in Eq.\ \ref{esym} with its existing constraint in the subsaturaion density region \cite{Tsang12}. As an example, shown in Fig.\ 2 is such a comparison with $S_0=30$ MeV. It is seen that for a given $\gamma$, depending on whether it is smaller or larger than $2/3$, the symmetry energy becomes softer or stiffer by reducing the value of $\eta$ (kinetic symmetry energy). For $S_0=30$ MeV, $\eta$ can be as small as zero. By varying the value of $S_0$ between 25 and 36 MeV covering the whole range currently used in the literature, we find that even negative kinetic symmetry energy is allowed, consistent with the information shown in Fig.1 and predictions in refs.\  \cite{XuLi,Vid11,Lov11,Car12,Rio14}. From Fig.\ \ref{figure2}, we also notice that the stiffness of the symmetry energy at suprasaturation densities is affected appreciably by  both the $\eta$ and $\gamma$ parameters, i.e., both the kinetic and potential parts of the symmetry energy.

The $E_{sym}(\rho)$ parameterized in Eq.\ \ref{esym} can be used directly to understand some experimental observables within statistical models when the thermal and chemical equilibrium have been reached. However, in heavy-ion collisions thermal equilibrium normally happens at the so-called freeze-out density below $\rho_0$. To extract nuclear symmetry energy at supra-saturation densities from heavy-ion collisions one has to use dynamical observables and understand well the role of the isospin degree of freedom during the reaction. This has been shown to be a very challenging task. To go one step further and get information about the separate kinetic and potential parts of the symmetry energy is much more difficult. How to explicitly incorporate properly SRC effects from the initialization of nucleons in phase space, to the in-medium elementary nucleon-nucleon collisions and the off-shell propagation of high-momentum nucleons during heavy-ion collisions is a complex problem on the agenda of our future work. In this exploratory study, we address a relatively simple question, namely,  under the condition that the symmetry energy $S_0$ at saturation density is fixed, how does the enhanced (reduced) potential ( kinetic) symmetry energy affect the isospin dynamics and isovector observables in heavy-ion collisions? The key to answer this question is the nucleon symmetry potential. Without considering the momentum dependence, the symmetry potential corresponding to the symmetry energy of Eq.\ \ref{esym} is
\begin{equation}\label{usym}
U^{n/p}_{\rm sym}(\rho,\delta)=[S_0-\eta\cdot E_{sym}^{kin}(\rho_0)(FG)]\cdot(\rho/\rho_0)^{\gamma}
\cdot[\pm 2\delta+(\gamma-1)\delta^2]
\end{equation}
where $\delta=(\rho_n-\rho_p)/\rho$ is the isospin asymmetry of the medium. We notice that the $\pm 2\delta$ term dominates and the $\pm$ sign is for $n/p$, namely, neutrons (protons) feel repulsive (attractive) symmetry potentials. Basically, the $\eta$ and $\gamma$ control respectively the magnitude and density dependence of the symmetry potential. While numerically varying the $\eta$ is equivalent to varying the $S_0$ as in some previous studies in the literature, they are conceptually different and have different consequences. We emphasize again that in our approach the $S_0$ is fixed at a value consistent with the existing experimental constraints by varying simultaneously and self-consistently the kinetic and potential parts of the symmetry energy in the opposite direction. The reaction dynamics is determined by the nuclear force, i.e., the density gradient of the potential, thus both the $\eta$ and $\gamma$ parameters affect the isospin dynamics. With $\eta=1$, the Eq.\ \ref{usym} reduces to the symmetry potential widely used by the heavy-ion reaction community especially in the earlier days, see, e.g. refs. \cite{Bar05,LCK} for a review.

\begin{figure}
\begin{center}
\includegraphics[width=17cm,height=16cm]{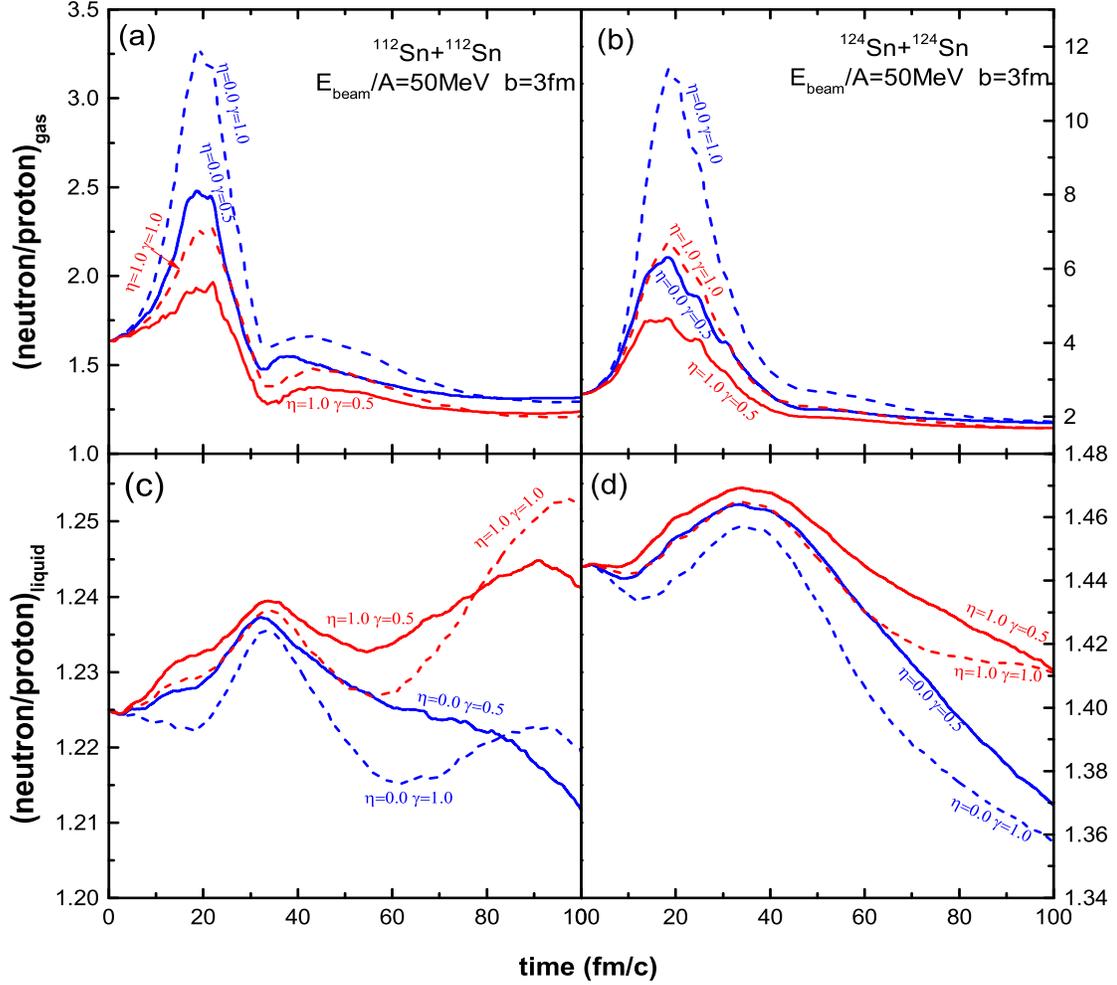}
\caption{(Color online) \label{np-time} Evolution of the neutron/proton ratios in the gas (liquid) regions in  $^{112}$Sn+$^{112}$Sn (left) and $^{124}$Sn+$^{124}$Sn (right) reactions at a beam energy of 50 MeV/nucleon and an impact parameter of  3 fm with different combinations of the kinetic and potential symmetry energies described in the text.}
\end{center}
\end{figure}
\section{Enhanced significance of the isospin degree of freedom in heavy-ion collisions with a
reduced kinetic symmetry energy}
\subsection{Evolution of the neutron/proton ratio and isospin fractionation}
Because of the generally increasing symmetry energy with density, one expects the low density region to become more neutron-rich compared to the denser regions simply from energy considerations. This is the so-called isospin fractionation. One can separate nucleons in the dilute/dense regions by using a cut on the nucleon local density.
Here we adopt a cutoff at $\rho_c=\rho_0/8$ widely used in the literature. Nucleons with $\rho\leq \rho_c$ are loosely described as in the gas phase while the rest are in the liquid phase. Of course, even in the initial state of the
reaction, nucleons near the surfaces of the colliding nuclei are also classified as in the gas phase. Shown in Fig.\ \ref{np-time} are the evolutions of the neutron/proton ratios in the gas (liquid) regions in $^{124}$Sn+$^{124}$Sn (left) and $^{112}$Sn+$^{112}$Sn (right) reactions at a beam energy of 50 MeV/nucleon and an impact parameter of  3 fm. To examine effects of the reduced (enhanced) kinetic (potential) symmetry energies, we compare results obtained with $\eta=1$ (with the free Fermi gas kinetic symmetry energy) and $\eta=0$ (no kinetic symmetry energy). We notice that it was shown that the kinetic symmetry energy at $\rho_0$ is actually reduced to about $-(9 \pm 7)$ MeV when the SRC is considered \cite{OrHen14}. Here we simply turn on or off the kinetic symmetry energy by setting $\eta=1$ or $0$ for illustrations. With both $\gamma=1$ or $0.5$, turning off the kinetic symmetry energy significantly enhances the degree of isospin fractionation making the gas phase more neutron-rich. Obviously, the effect is stronger for the more neutron-rich reaction system of $^{124}$Sn+$^{124}$Sn. Since the EOS and symmetry potential depend on the isospin asymmetry $\delta$ quadratically and basically linearly, respectively, the enhanced isospin fractionation will subsequently influence the isospin dynamics and isovector observables.

\begin{figure}
\begin{center}
\includegraphics[width=14cm,height=18cm]{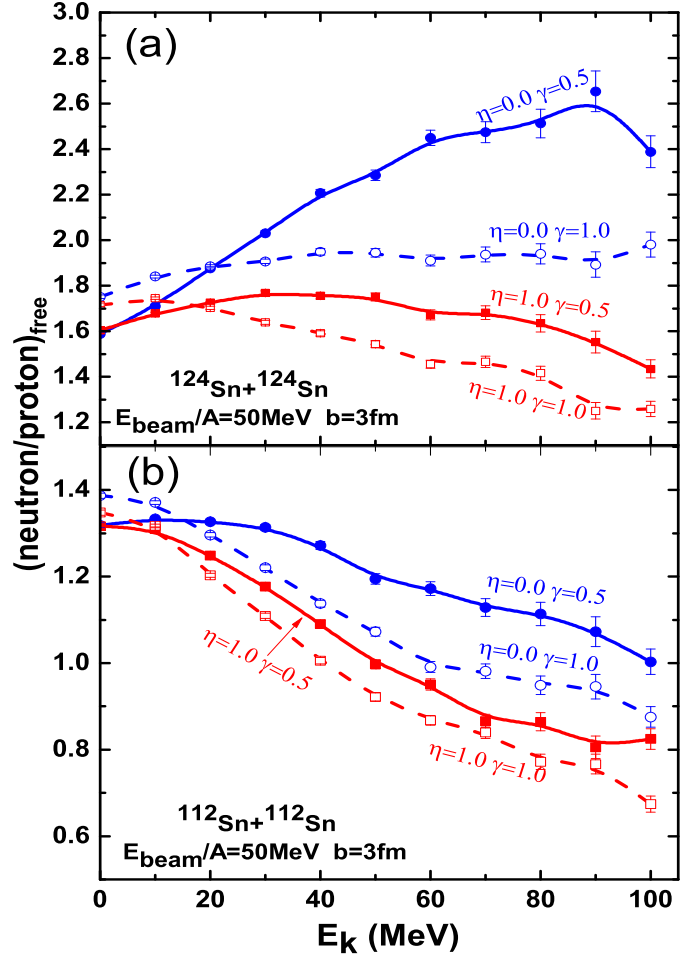}
\caption{(Color online) \label{npek50} The free neutron/proton ratio as a function of nucleon kinetic energy in $^{124}$Sn+$^{124}$Sn (upper panel) and $^{112}$Sn+$^{112}$Sn (lower panel) reactions at a beam energy of 50 MeV/nucleon and an impact parameter of  3 fm with different combinations of the kinetic and potential symmetry energies described in the text.}
\end{center}
\end{figure}

\begin{figure}
\begin{center}
\includegraphics[width=14cm,height=18cm]{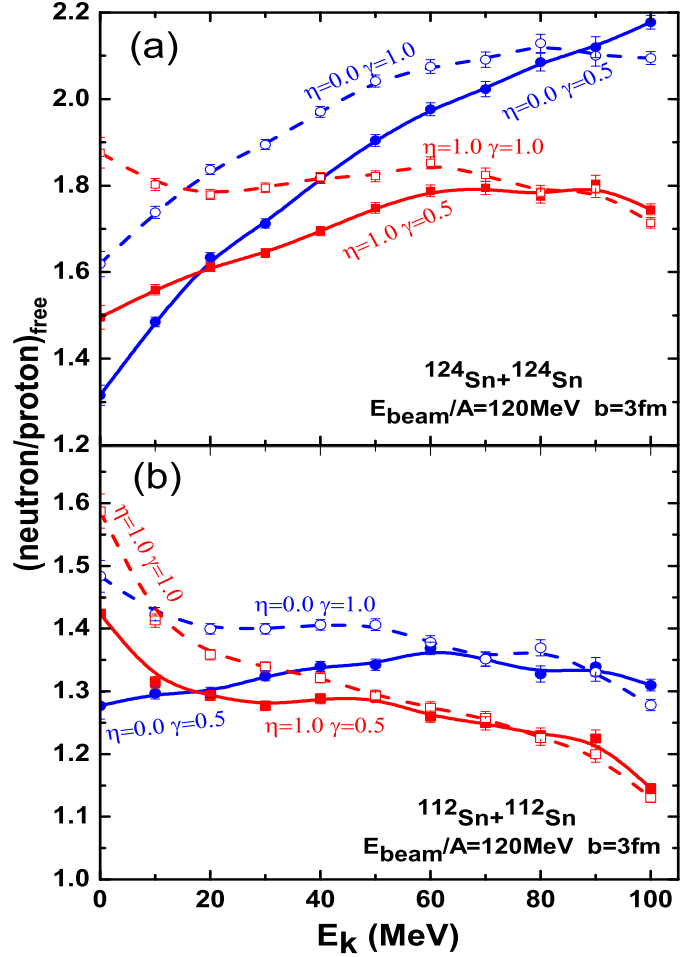}
\caption{(Color online) \label{npek120} Same as in Fig. \ref{npek50} but at a beam energy of 120 MeV/nucleon.}
\end{center}
\end{figure}

\begin{figure}
\begin{center}
\includegraphics[width=12cm,height=12cm]{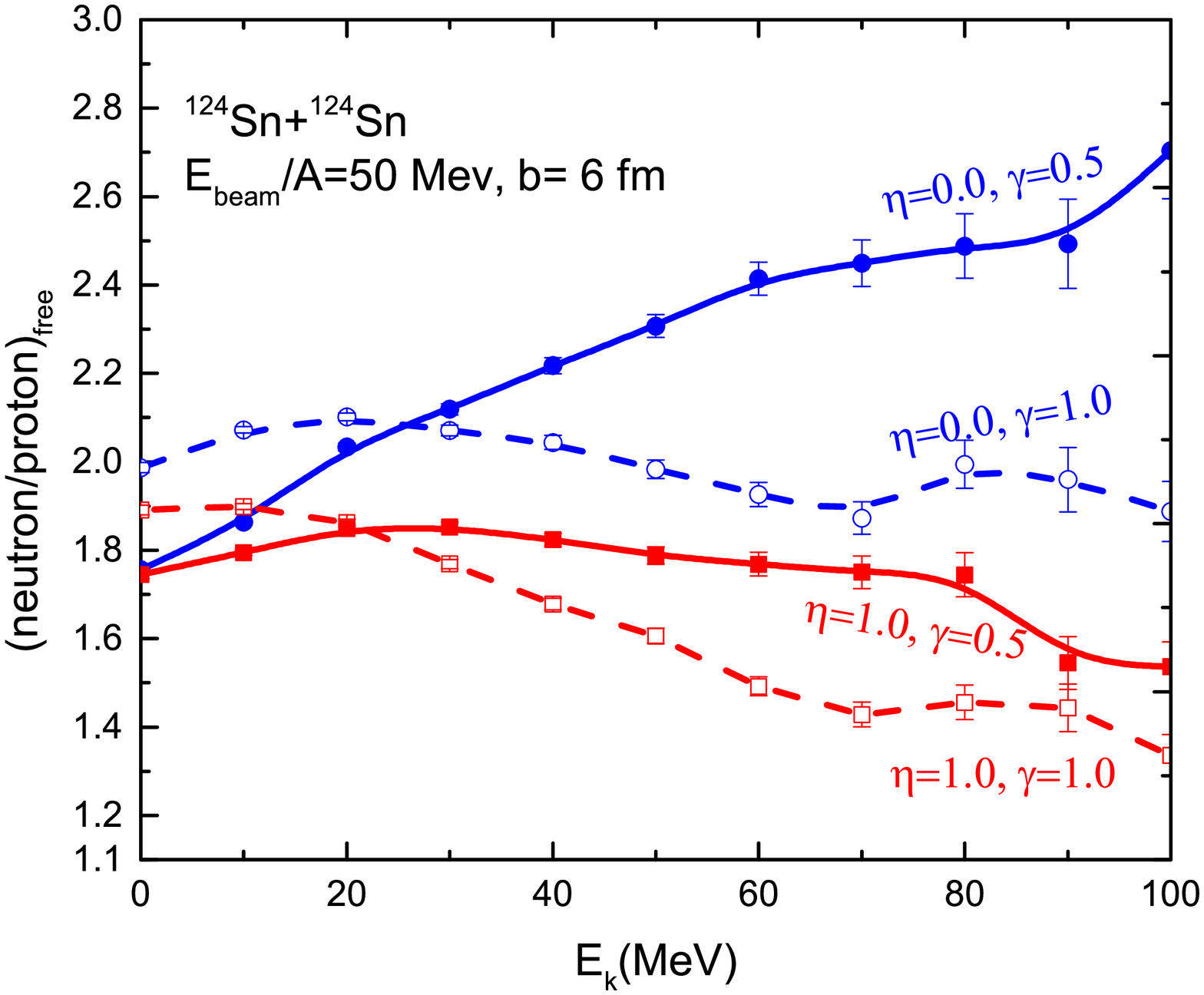}
\caption{(Color online) \label{npek50b6}
Same as in window (a) of Fig. \ref{npek50} but at an impact parameter of 6 fm.}
\end{center}
\end{figure}

\begin{figure}[htb]
\begin{center}
\includegraphics[width=12cm,height=12cm]{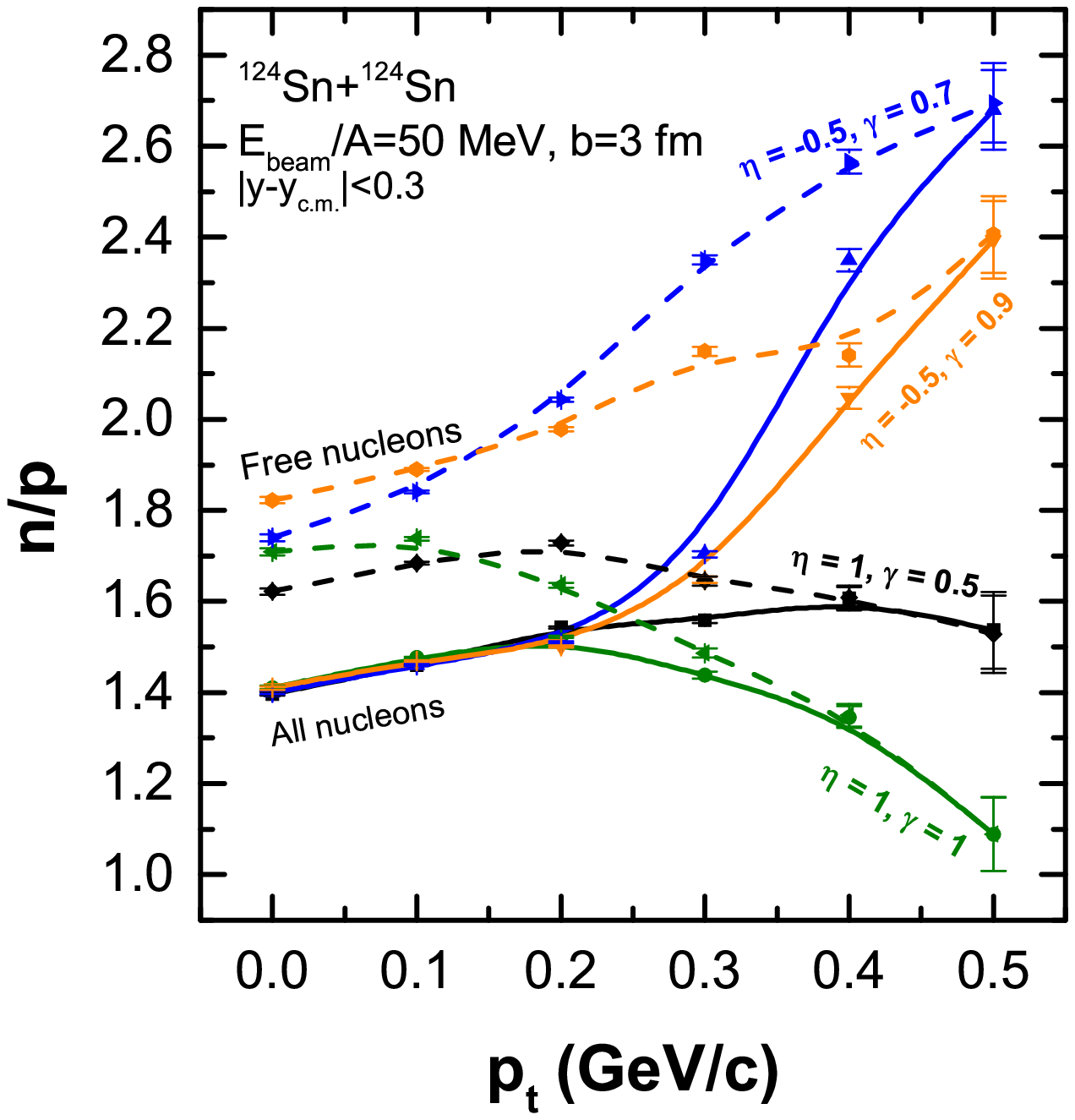}
\caption{(Color online) \label{nppt} Transverse momentum dependence of the neutron/proton ratio at mid-rapidity in $^{124}$Sn+$^{124}$Sn reactions at a beam energy of 50 MeV/nucleon and an impact parameter of  3 fm.}
\end{center}
\end{figure}
\subsection{Free neutron/proton ratio at freeze-out}
At the freeze-out, the neutron/proton ratio of the gas phase naturally becomes the free neutron/proton ratio experimentally measurable. Shown in Fig.\ \ref{npek50} and Fig.\ \ref{npek120} are the
free neutron/proton ratio as a function of nucleon kinetic energy in $^{124}$Sn+$^{124}$Sn (upper panel) and $^{112}$Sn+$^{112}$Sn (lower panel) reactions at an impact parameter of  3 fm
and a beam energy of 50 MeV/nucleon and 120 MeV/nucleon, respectively. As one expects, the free neutron/proton ratio depends on both the $\eta$ and $\gamma$ parameters.
It is seen that calculations at 50 MeV/nucleon with $\eta=0$ lead to significantly higher free neutron/proton ratios especially for more energetic nucleons as they are mostly from the earlier stage of the reaction where the density is higher.  At this beam energy, the maximum density reached is only about $1.2\rho_0$ in the central region. Most of the particles are actually in the subsaturation density regions during the entire reaction process. As shown in Fig.\ \ref{figure2}, in the subsaturation density region the symmetry energy with $\gamma=0.5$ is higher than that with $\gamma=1$, while it is the opposite at suprasaturation densities. One can thus easily understand the feature shown in Fig.\ \ref{npek50} that $\gamma=0.5$ leads to higher free neutron/proton ratios than $\gamma=1.0$ for a given $\eta$.

As the beam energy increases to 120 MeV/nucleon, some interesting changes occur. First of all, the low energy nucleons are now more sensitive to both the $\eta$ and $\gamma$ parameters. The energetic nucleons are now mainly affected by the variation of $\eta$ especially in the $^{112}$Sn+$^{112}$Sn reaction. At this higher beam energy, the maximum density reachable is about $1.7-2\rho_0$. It has been known that the free neutron/proton ratio in reactions with beam energies far above the Fermi energy becomes less sensitive to the density dependence of the symmetry energy when nucleon-nucleon collisions dominate over the mean-field in the reaction dynamics and the ratio of isovector/isoscalar potential becomes smaller at higher densities.  The free neutron/proton ratio of low energy nucleons is still affected by the variation of both $\eta$ and $\gamma$.  It is interesting to see that at $E_{beam}/A=120$ MeV, stiffer symmetry energy with $\gamma=1$ leads to a higher free neutron/proton ratio for a given $\eta$ in contrast to the case of $E_{beam}/A=50$ MeV. This is because of the different densities reached in the two cases and the cross of the symmetry energy from below to above $\rho_0$ with different $\gamma$ parameters for a given $\eta$.

To examine effects of the impact parameter, we show in Fig.\ \ref{npek50b6} the free neutron/proton ratio as a function of nucleon kinetic energy in the $^{124}$Sn+$^{124}$Sn reaction at a beam energy of 50 MeV/nucleon and an impact parameter of 6 fm. Comparing with results of the same reaction but at an impact parameter of 3 fm shown in the window (a) of Fig.\ \ref{npek50}, we see that effects of the reduced kinetic symmetry energy are qualitatively the same. Of course, with the same number of events the statistics becomes poor especially at high nucleon kinetic energies in the more peripheral reactions.

Mid-rapidity nucleons are mostly from the participant regions of heavy-ion collisions. They may thus show higher sensitivity to the symmetry energy. As an example, shown in Fig.\ \ref{nppt} are the neutron/proton ratios as a function of nucleon transverse momentum in the $^{124}$Sn+$^{124}$Sn
reaction at 50 MeV/nucleon. The free neutron/proton ratio is higher than that for all nucleons (including bounded ones) as one expects. At high transverse momenta, all nucleons are free and they indeed show a larger sensitivity to the variation of both $\eta$ and $\gamma$. In experiments, to reduce the systematic errors associated with the measurement of neutrons, one sometimes takes the double ratio of the free neutron/proton in two reactions. We found, however, the double ratio for the two Sn+Sn reactions considered here significantly reduces the sensitivity to both the $\eta$ and $\gamma$ parameters compared to the single neutron/proton ratio especially at higher beam energies.

\begin{figure}[htb]
\begin{center}
\includegraphics[width=12cm,height=14cm]{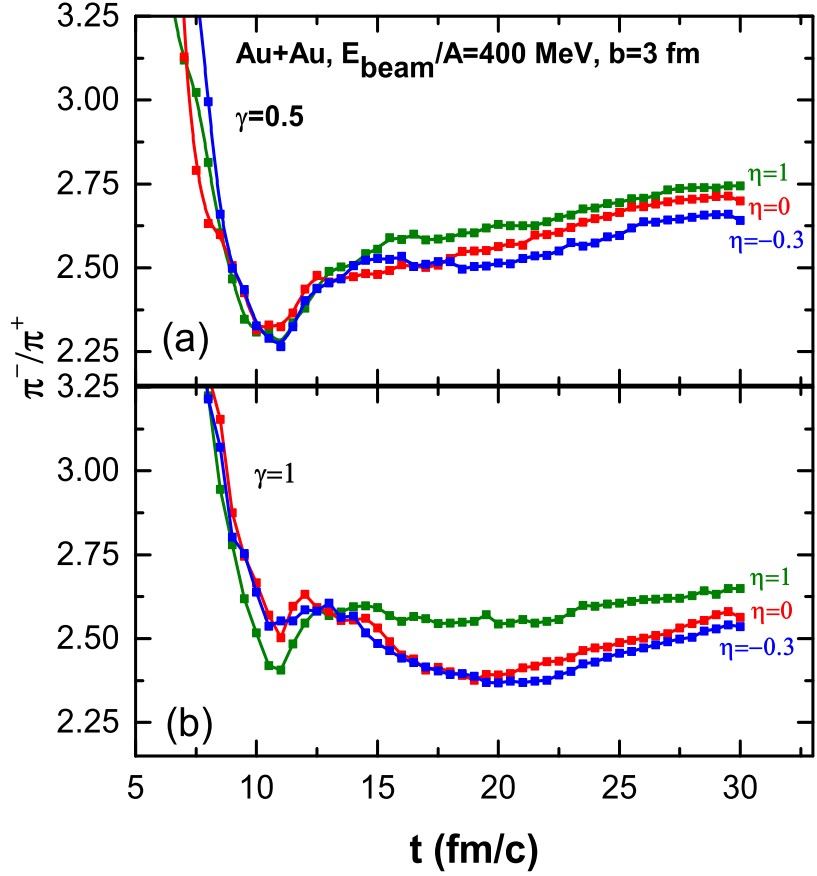}
\caption{(Color online) \label{pion} Evolution of the $\pi^-/\pi^+$ ratio in Au+Au reaction at a beam energy of 400 MeV/nucleon and an impact parameter of 3 fm.}
\end{center}
\end{figure}

\subsection{Evolution of the $\pi^-/\pi^+$ ratio}
At beam energies above the pion production threshold, besides the neutron/proton ratio the $\pi^-/\pi^+$  ratio is another isospin tracer and it has been known as a sensitive probe of the high-density behavior of nuclear symmetry energy \cite{Li02-pion}. It is interesting to know how the reduced kinetic symmetry energy may affect the evolution of the $\pi^-/\pi^+$  ratio. Shown in Fig.\ \ref{pion} are the $\pi^-/\pi^+$  ratio in Au+Au reactions at a beam energy of 400 MeV/nucleon and an impact parameter of 3 fm. First of all, consistent with what is known before, the softer ($\gamma=0.5$) symmetry energy predicts a higher $\pi^-/\pi^+$  ratio at freeze-out. Reducing the kinetic symmetry energy from the free Fermi gas prediction ($\eta=1$) decreases the final $\pi^-/\pi^+$  ratio. Earlier studies have indicated that the $\pi^-/\pi^+$  ratio reflects the neutron/proton ratio of the high density region \cite{Mzhang}.
Regardless of the value of $\gamma$, reducing $\eta$ makes the neutron/proton ratio of the high density region higher as shown by the (neutron/proton)$_{\text{liquid}}$ in the lower panels of Fig.\ \ref{np-time}. Thus, the $\pi^-/\pi^+$  ratio is higher with decreasing $\eta$. Overall, effects of the $\eta$ and $\gamma$ are comparable.

\section{Summary and discussions}
In summary, there are solid theoretical basis and experimental evidence that the short-range nucleon-nucleon correlation reduces the kinetic symmetry energy significantly compared to the free Fermi gas model prediction. In this work, we have shown that existing constraints on the density dependence of nuclear symmetry energy around saturation density do not limit the partition of kinetic and potential parts of the symmetry energy. Current constraints on the total symmetry energy can accommodate the reduced (correspondingly enhanced) kinetic (potential) symmetry energy in a broad range. Fixing the total symmetry energy at saturation density at a constant consistent with the current constraints available, the reduced (enhanced) kinetic (potential) symmetry energy strengthens significantly the role played by the isospin degree of freedom in heavy-ion collisions. Some experimental consequences are discussed. In particular, the evolution of the neutron/proton and $\pi^-/\pi^+$ ratio as well as the kinetic energy and transverse momentum dependence of the free neutron/proton ratio at the freeze-out of heavy-ion collisions are all strongly affected by the reduced kinetic symmetry energy due to the short-range nucleon-nucleon correlation.

We would like to re-emphasize that the main purpose of this exploratory work is to get a qualitatively understanding of the effects of the SRC reduced kinetic symmetry energy in heavy-ion collisions. As we mentioned earlier, a lot more work needs to be done to draw a strong conclusion from comparing with data quantitatively. In particular, to incorporate consistently SRC effects in the initialization of colliding nuclei, off-shell propagation of high-momentum nucleons and the momentum-dependence of the symmetry potential in transport models remains an interesting challenge.
Nevertheless, it is worth noting that current indications for a reduced kinetic symmetry energy are strong. For example, without considering the SRC reduced kinetic symmetry energy, IBUU calculations \cite{Kon15} fall far below the NSCL/MSU data on the free neutron/proton double ratio from central $^{124}$Sn+$^{124}$Sn and $^{112}$Sn+$^{112}$Sn collisions at 50 and 120 MeV/u~\cite{Cou14}. This failure calls for new mechanisms to enhance the double neutron/proton ratio. Interestingly, with all the cautions mentioned above, calculations using the same model as in the present work can well reproduce the NSCL/MSU data \cite{OrHen14}. In fact, by performing the $\chi^2$ fit to the NSCL/MSU data in the $\eta-\gamma$ parameter plane, we found that the best combination is $\eta=-0.30(1\pm 18.53\%)$ and $\gamma=0.80(1\pm 5.98\%)$, corresponding to a kinetic symmetry energy of $E_{sym}^{kin}(\rho_0)=-(3.8\pm 0.7)$ MeV at $\rho_0$ \cite{OrHen14}. We thus conclude that effects of the SRC reduced kinetic symmetry energy in heavy-ion collisions at intermediate energies should be considered seriously.

\section{Acknowledgement}
We would like to thank F.J. Fattoyev, X.T. He, O. Hen, X. H. Li, E. Piasetzky,  L.B. Weinstein and W.G. Newton for helpful discussions. This work is supported in part by  the US National Science Foundation under Grant No. PHY-1068022 and PHY-1359409 (REU), US National Aeronautics and Space Administration under Grant No. NNX11AC41G issued through the Science Mission Directorate, the CUSTIPEN (China-U.S. Theory Institute for Physics with Exotic Nuclei) under DOE grant number DE-FG02-13ER42025, the National Natural Science Foundation of China under Grant No. 11320101004 and 10905041, and the China Scholarship Council Foundation (201208310156).


\begin{thebibliography}{99}

\bibitem{EPJA} B.A. Li, A. Ramos, G. Verde, and I. Vida\~na, eds., "Topical issue on nuclear symmetry energy", Eur. Phys. J. A {\bf 50}, No. 2, (2014).

\bibitem{Bom91} I. Bombaci, U. Lombardo, Phys. Rev. C 44 (1991) 1892.

\bibitem{XuLi} C. Xu and B.A. Li, arXiv: 1104.2075; C. Xu, A. Li, B.A. Li, J. of Phys: Conference Series 420,  012190 (2013).

\bibitem{Vid11} I. Vidana, A. Polls, C. Providencia, Phys Rev C 84, 062801(R) (2011).

\bibitem{Lov11} A. Lovato, O. Benhar, S. Fantoni, A. Yu. Illarionov, and K. E. Schmidt, Phys. Rev. C 83, 054003 (2011).

\bibitem{Car12} A. Carbone, A. Polls, A. Rios, Eur. Phys. Lett. 97,  22001 (2012).

\bibitem{Rio14} A. Rios, A. Polls, W. H. Dickhoff, Phys. Rev. C 89, 044303 (2014).

\bibitem{Hen14} O. Hen \textit{et al.}, Science \textrm{346}, 614 (2014).

\bibitem{OrHen14} Or Hen, Bao-An Li, Wen-Jun Guo, L.B. Weinstein, Eli Piasetzky,
Phys. Rev. C \textbf{91}, 025803 (2015).

\bibitem{LiBA13} B.A. Li, X. Han, Phys. Lett. B \textbf{727}, 276 (2013).

\bibitem{LCK} B. A. Li, L. W. Chen, and C. M. Ko, Phys. Rep. \textbf{464}, 113 (2008).

\bibitem{Bar05} V. Baran, M. Colonna, V. Greco, and M. Di Toro, Phys. Rep. \textbf{410},335 (2005).

\bibitem{Lynch09} W.G. Lynch {\it et al.}, Prog. Nucl. Part. Phys. {\bf 62}, 427 (2009).

\bibitem{Trau12} W. Trautmann and H. H. Wolter, Int. J. Mod. Phys. E {\bf 21}, 1230003 (2012).

\bibitem{Tsang12} M. B. Tsang, et al., Phys. Rev. C {\bf 86}, 015803 (2012).

\bibitem{Hor14} C.J. Horowitz et al., J. of Phys. G {\bf 41}, 093001 (2014).

\bibitem{Jim13} J. M. Lattimer, Annu. Rev. Nucl. Part. Sci. \textbf{62}, 485 (2012).

\bibitem{lat12} James M. Lattimer,  Annu. Rev. Nucl. Part. Sci. 62, 485 (2012).

\bibitem{Tamii} A. Tamii, P. vonNeumann-Cosel and I. Poltoratska, Euro. Phys. Jour. A 50:28 (2014).

\bibitem{Rios}R. Sellahewa and A. Rios, Phys. Rev. C{\bf 90}, 054327 (2014).

\bibitem{Li02-pion} B.A. Li, Phys. Rev. Lett. 88, 192701 (2002).

\bibitem{Mzhang} M. Zhang, Z.G. Xiao, B.A. Li, L.W. Chen, G.C. Yong and S.J. Zhu, Phys. Rev. C{\bf 80}, 034616 (2009).

\bibitem{Kon15} H.Y. Kong, Y. Xia, J. Xu, L.W. Chen, B.A. Li, and Y.G. Ma, arXiv:1502.00778, Phys. Rev. C (2015) in press.
\bibitem{Cou14} D.S. Coupland et al., arXiv:1406.4546

\end{thebibliography}
\end{document}